\documentstyle[11pt,newpasp,twoside,epsf]{article}
\newcommand{\beq}       {\begin{equation}}
\newcommand{\eeq}       {\end{equation}}
\newcommand{\beqa}      {\begin{eqnarray}}
\newcommand{\eeqa}      {\end{eqnarray}}

\def\avg#1              {\langle #1\rangle}     

\newcommand{\calc}      {\ifmmode {{\cal C}} \else ${{\cal C}}$\fi}
\newcommand{\calf}      {\ifmmode {{\cal F}} \else ${{\cal F}}$\fi}
\newcommand{\calg}      {\ifmmode {{\cal G}} \else ${{\cal G}}$\fi}
\newcommand{\calh}      {\ifmmode {{\cal H}} \else ${{\cal H}}$\fi}
\newcommand{\call}      {\ifmmode {{\cal L}} \else ${{\cal L}}$\fi}
\newcommand{\calm}      {\ifmmode {{\cal M}} \else ${{\cal M}}$\fi}
\newcommand{\caln}      {\ifmmode {{\cal N}} \else ${{\cal N}}$\fi}
\newcommand{\calo}      {\ifmmode {{\cal O}} \else ${{\cal O}}$\fi}
\newcommand{\calp}      {\ifmmode {{\cal P}} \else ${{\cal P}}$\fi}
\newcommand{\calq}      {\ifmmode {{\cal Q}} \else ${{\cal Q}}$\fi}
\newcommand{\calr}      {\ifmmode {{\cal R}} \else ${{\cal R}}$\fi}
\newcommand{\cals}      {\ifmmode {{\cal S}} \else ${{\cal S}}$\fi}
\newcommand{\calt}      {\ifmmode {{\cal T}} \else ${{\cal T}}$\fi}
\newcommand{\calv}      {\ifmmode {{\cal V}} \else ${{\cal V}}$\fi}
\newcommand{\calw}      {\ifmmode {{\cal W}} \else ${{\cal W}}$\fi}

\newcommand{\cm}        {\ifmmode {\rm cm}\else cm\fi}
\newcommand{\m}         {\ifmmode {\rm m}\else m\fi}
\newcommand{\km}        {\ifmmode {\rm km}\else km\fi}
\newcommand{\pc}        {\ifmmode {\rm pc}\else pc\fi}
\newcommand{\kpc}       {\ifmmode {\rm kpc}\else kpc\fi}
\newcommand{\Mpc}       {\ifmmode {\rm Mpc}\else Mpc\fi}
\newcommand{\Gpc}       {\ifmmode {\rm Gpc}\else Gpc\fi}
\newcommand{\ly}        {\ifmmode {\rm ly}\else ly\fi}
\newcommand{\Rsun}      {\ifmmode {R_\odot}\else R$_\odot$ \fi}
\newcommand{\AU}        {\ifmmode {{\rm AU}}\else AU \fi} 
\newcommand{\s}         {\ifmmode {\rm s}\else s\fi}
\newcommand{\Hz}        {\ifmmode {\rm Hz}\else Hz\fi}
\newcommand{\yr}        {\ifmmode {\rm yr}\else y\fi}
\newcommand{\cms}       {\ifmmode {\rm cm~s^{-1}}\else cm~s$^{-1}$\fi}
\newcommand{\kms}       {\ifmmode {\rm km~s^{-1}}\else km~s$^{-1}$\fi}
\newcommand{\K}         {\ifmmode {\rm K}\else K\fi}
\newcommand{\ster}      {\ifmmode {\rm ster}\else ster\fi}
\newcommand{\erg}       {\ifmmode {\rm erg}\else erg\fi}
\newcommand{\dyn}       {\ifmmode {\rm dyn}\else dyn\fi}
\newcommand{\mug}       {\ifmmode {\mu \rm{G}} \else {$\mu$G}\fi}
\newcommand{\Msun}      {\ifmmode {M}_{\mathord\odot}\else 
                          $M_{\mathord\odot}$\fi}
\newcommand{\msun}      {\ifmmode {M}_{\mathord\odot}\else 
                          $M_{\mathord\odot}$\fi}
\newcommand{\Lsun}      {\ifmmode {L}_{\mathord\odot}\else 
                          $L_{\mathord\odot}$\fi}
\newcommand{\rate}      {\ifmmode {\rm cm^3~s^{-1}}\else cm$^3$~s$^{-1}$\fi}

\newcommand{\e}         {\ifmmode ^{-1}\else $^{-1}$\fi}
\newcommand{\ee}        {\ifmmode ^{-2}\else $^{-2}$\fi}
\newcommand{\eee}       {\ifmmode ^{-3}\else $^{-3}$\fi}

\newcommand{\mdw}       {\dot m_w}
\newcommand{\mds}       {\dot m_*}

\newcommand{\smyr}{{ M_\odot\ \rm yr^{-1}}}
\newcommand{\sm}{{ M_\odot}}

\def\ion#1#2{#1$\;${\small\rm II}\relax}
\def\lesssim{\mathrel{\hbox{\rlap{\hbox{\lower4pt\hbox{$\sim$}}}\hbox{$<$}}}}
\def\gtrsim{\mathrel{\hbox{\rlap{\hbox{\lower4pt\hbox{$\sim$}}}\hbox{$>$}}}}

\def\edcomment#1{\iffalse\marginpar{\raggedright\sl#1\/}\else\relax\fi}
\reversemarginpar

\begin{document}
\vspace{-0.2in}
\title{Testing Massive Star Formation Theory in Orion}
\vspace{-0.2in}
\author{Jonathan C. Tan}
\affil{Princeton University Observatory, Princeton, NJ 08544, USA}

\vspace{-0.1in}
\begin{abstract}
  I compare theoretical models of massive star formation with
  observations of the Orion Hot Core, which harbors one of the closest
  massive protostars. Although this region is
  complicated, many of its features (size, luminosity, accretion disk,
  \ion{H}{2} region, outflow) may be understood by starting with a
  simple model in which the star forms from a massive gas
  core that is a coherent entity in approximate pressure balance with
  its surroundings.  The dominant contribution to the pressure is from
  turbulent motions and magnetic fields. The collapse can be perturbed
  by interactions with other stars in the forming cluster, which may
  induce sporadic enhancements of the accretion and outflow rate.
\vspace{-0.2in}
\end{abstract}
\vspace{-0.2in}
\section{Theoretical Models of Massive Star Formation}
\vspace{-0.1in}
Stars much more massive than the Sun, although rare, are
important for the energetics and metal production of galaxies. How do
these stars assemble themselves from the interstellar medium (ISM)?

Observationally it is clear that massive stars are born in the densest
{\it clumps} of gas inside giant molecular clouds (GMCs) (e.g.
Mueller et al. 2002), which undergo quite efficient ($\sim 10 - 50\%$)
transformation to star clusters. The new stellar mass is mostly in
low-mass stars, and it appears that the majority of Galactic star
formation occurs in this clustered mode (Lada \& Lada 2003).  This
concentration of star formation in a relatively small part of the
total Galactic molecular ISM, suggests that the
creation of clumps may be triggered by processes external to GMCs. One
possibility is an origin in local regions of pressure enhancement
created in GMC collisions (Tan 2000). An alternative to triggering is
the gradual condensation of clumps in regions of GMCs that become
sufficiently self-shielded from the Galactic far UV background (McKee
1989).

This question can be addressed by studying the infrared dark clouds
(IDCs) (e.g. Egan et al. 1998),
which are the likely precursors of star-forming clumps.  The
collisional model predicts IDCs are surrounded by coherent, supersonic
($\sim 10\kms$) flows. Teyssier et al. (2002) report significant
velocity structure towards all their IDCs, and in at least one case
the gas is spatially connected across this velocity range. Since the
angular momentum vectors of collisions in a thin shearing disk can be
both parallel and anti-parallel to that of the host galaxy, the
collisional model can account for the almost equal proportions of pro-
and retrograde GMC rotations in M33 (Rosolowsky et al. 2003).
The dependence of collision rate on shear leads to reduced star
formation efficiency in galaxies with rising rotation curves ---
a general trend of the Hubble sequence.

In this article I focus on the separate question of how clumps, once
formed, transform a small part of themselves into massive stars.  Clumps
can be regarded as quasi-virialized structures: virial mass estimates
are similar to estimates of the total gas and stellar mass (Plume et
al. 1997); and their morphologies are often close to spherical
(Shirley et al. 2003). The virial velocity is typically several $\kms$,
while strong cooling to $\sim10$~K causes the sound speed to be
only $c_{\rm th}= 0.19(T/10{\rm K})^{1/2}\kms$ (for $n_{\rm
  He}=0.2n_{\rm H_2}$). Thus the clumps are supersonically turbulent.
Measured magnetic field strengths are $\sim {\rm mG}$ (Crutcher \& Lai
2002).
The Alfv\'en velocity, $v_A=B/(4\pi \rho_0)^{1/2}=1.84(B/{\rm mG})(n_{\rm
  H}/10^6{\rm cm^{-3}})^{-1/2}\kms$, is comparable to the virial velocity.

Turbulence and self-gravity engender the clumps with substructure, the
nature of which is under intense numerical study (see Mac-Low \&
Klessen 2003 for a review). Self-gravity tends to cause condensations
on the scale of the Jeans mass, which can be generalized to include
nonthermal forms of pressure support.  Turbulence leads to compression
of gas into filaments and sheets.  The combination may be enough to
produce the observed mass spectrum of cores, which appears to be
similar to the stellar mass function (e.g. Motte et al. 2001).
Massive, quiescent cores are seen in Orion (e.g. Li, Goldsmith, \&
Menten 2003).

\vspace{-0.06in}
\subsection{Formation from Direct Collapse of Gas Cores}
\vspace{-0.05in}

The rate of collapse of gravitationally unstable gas cores is set by
their initial density profile. Singular isothermal spheres, described
by the Shu solution for inside-out collapse, have a constant accretion
rate, $\mds=0.975 c_{\rm th}^3/G=1.54\times 10^{-6} (T/10{\rm
  K})^{3/2}\smyr$, which is directly related to the local enclosed
mass divided by the local free-fall timescale: $\mds=\phi_* m_*/t_{\rm
  ff}$, with $\phi_*=0.663$. The collapse of uniform spheres,
described by the Larson-Penston solution, has a value of $\phi_*$
that is about 50 times larger. Bonnor-Ebert spheres have non-singular
centers, that locally approximate uniform density, so their
initial collapse rate is relatively high, but then evolves towards the
Shu solution. For pressure-confined clouds, as the pressure
is raised, the size of the equilibrium cloud contracts, thus
increasing its mean density and accretion rate. The above results
apply in a similar manner for the more general case of cores with
polytropic equations of state. Differences in the density profile with
respect to the $r^{-2}$ law of the singular isothermal sphere, lead to
departures from a constant accretion rate: e.g. a shallower
profile leads to a growing accretion rate.

Myers \& Fuller (1992) and McLauglin \& Pudritz (1997) considered
various models for the density structure of massive star-forming
cores, but normalized them to be in equilibrium in a medium of quite
low pressure.
Formation timescales could then be $\gtrsim 10^6\:{\rm yr}$, a
significant fraction of the main sequence stellar lifetime and
probably inconsistent with the relative small spread in ages seen in
young star clusters (Palla \& Stahler 1999). Osorio, Lisano, \&
D'Alessio (1999) considered collapse with higher accretion rates, but
with the normalization set via an empirical modeling of the infrared
spectra of the gas envelopes.

McKee \& Tan (2003, hereafter MT) approximated the structure of massive gas
cores that are about to undergo collapse to stars with
pressure-truncated singular polytropic spheres, including the effect
of a thermal core. The particular equation of state
($P=K\rho^{\gamma_p}$, with $\gamma_p=2/3$) was chosen so that the
equilibrium density profile matched observed profiles of {\it clumps}
($\rho \propto r^{-1.5}$), and it was assumed that the same density
structure applied on the smaller scales of cores. The cores are bounded
by the mean pressure in the clump, which is estimated assuming clumps
are in approximate hydrostatic and virial equilibrium, so that
$P\simeq G\Sigma^2$, where $\Sigma$ is the surface density.
Typical observed values are $\Sigma\sim 1\:{\rm g\:cm^{-2}}$ for
clumps forming massive stars (Mueller et al. 2002).

Now consider the properties of a core of a given mass, $M$, that will
soon collapse to form a star. The equilibrium radius is
$r_c=0.057 M_{60}^{1/2} \Sigma^{-1/2}\:{\rm pc}$,
where $M_{60}=M/60\sm$.  
Cores are very concentrated, which alleviates the crowding
problem of formation in stellar clusters: the central
stellar density in the Orion Nebula Cluster (ONC) is
$\sim10^4\:{\rm pc^{-3}}$, giving a stellar separation of about
$0.05\:{\rm pc}$.

Note that the assumption the core collapse starts in the equilibrium state is
an idealization. Cores probably form and become unstable at the
confluence of turbulent flows, and so may be
somewhat out of equilibrium. However we expect that the deviations should generally be rather modest, as the turbulent motions are not
much greater than the Alfv\'en speed and as magnetic fields, both
tangled and ordered, are important sources of pressure support.

The velocity dispersion at the core surface is $1.27 M_{60}^{1/4}
\Sigma^{1/4}\kms$.  The minimum equilibrium core mass, the
Bonnor-Ebert mass where thermal pressure dominates, can be estimated
by setting this speed equal to the sound speed. This mass is $0.0504
(T/20{\rm K})^2 \Sigma^{-1}\sm$, which is comparable to the mass at which
the ONC mass function rapidly decreases (Muench et al. 2002).

The rate of core collapse is $\mds = 4.6\times 10^{-4}
f_*^{1/2} M_{60}^{3/4}\Sigma^{3/4}\:\smyr$, where $f_*$ is the ratio
of $m_*$ to the final stellar mass and 50\% formation efficiency is assumed.
Feeding a star and disk at such high rates may strongly influence the
star formation process. For example, in the limit of spherical
accretion, Wolfire \& Cassinelli (1987) pointed out that the ram
pressure of infalling gas at the dust destruction front could overcome
radiation pressure from a high-mass star.

The collapse time, $1.3\times 10^5 M_{60}^{1/4} \Sigma^{-3/4}\:{\rm
  yr}$, is short and quite insensitive to $M$. This allows
coeval star formation in clusters, consistent, for example, with the
estimated 1~Myr formation timescale of the ONC (Palla
\& Stahler 1999).

If the core starts with a rotational to gravitational energy ratio
$\beta$, then a disk forms at a centrifugal radius of about $r_{\rm
  disk}=1200 (\beta/0.02) (f_* M_{60})^{1/2} \Sigma^{-1/2}{\rm AU}$,
assuming solid body rotation of the core.  We have normalized to a typical value
of $\beta$ inferred in cores of lower mass and density (Goodman et al.
1993). 

Disk accretion is expected to be accompanied by an outflow of material
at a rate $\mdw\equiv f_w \mds$, with $f_w\simeq 0.1-0.4$, and a
velocity $v_w = f_v v_K = 920
(f_v/2.1)(m_*/10\sm)^{1/2}(r_*/10\:R_\odot)^{-1/2}\kms$, with e.g.
$f_v\simeq2.1$ (Shu et al. 2000), and where $v_K$ is the Keplerian
speed at the star.
A bipolar outflow is created perpendicular to the disk, which should
maintain its orientation over much of the star formation timescale,
unless the disk is perturbed by a companion, passing star, or warping
instability. Note that if many stars are forming together in a
cluster, then multiple outflows are inevitable (Tan \& McKee 2002), and
their effects must be disentangled.


Thus key signatures of this star formation model are the presence of
coherent gas cores, that contain protostellar disks, from which outflows
are driven and maintained for many local dynamical timescales. Our (Tan \&
McKee) recent research program has focused on trying to quantify the
properties of these elements of the model for comparison to
observations.

\vspace{-0.1in}
\subsection{Formation via Competitive Accretion and/or Stellar Collisions}

A number of objections to the core accretion model have been raised.
Theoretically, there is the problem of radiation pressure preventing
accretion to a massive, luminous protostar (Larson \& Starrfield 1971;
Wolfire \& Cassinelli 1987). A disk geometry may help (e.g. Nakano
1989; Yorke \& Sonnhalter 2002).  Observationally, it appears that
massive stars tend to form in crowded regions near cluster centers
(Bonnell, \& Davies 1998) and in binaries where the secondary is
relatively massive compared to a random sampling from the IMF
(Eggleton, Tout, \& Fitchett 1989).  Relatively large numbers of
massive stars are ``runaways'', perhaps ejected from dynamical
interactions in young star clusters (Gies 1987).

These points have motivated formation models based on protostellar
collisions and competitive Bondi-Hoyle accretion.  Bonnell, Bate, \&
Zinnecker (1998) presented a model in which extreme stellar densities
result in a cluster of lower-mass stars that dissipate their kinetic
energy as they accrete from the initially dominant gaseous component
of the protocluster. For the collisional timescale to become short
enough to be relevant to the formation process the stellar density
must reach at least $\sim 10^6 - 10^8\:{\rm pc^{-3}}$, several orders
of magnitudes greater than the observed central density of the ONC.
Bonnell \& Bate (2002) presented SPH simulations of the collapse of an
isothermal gas clump initially seeded with many low-mass stars. With a
collision radius of 2~AU, they found that the most massive star that
formed did increase its mass significantly in several merger events.
However, the large increase in density after one clump free-fall time,
when much of the growth occurs, is probably an artifact of the initial
conditions: i.e. the cold, synchronized collapse of a system that has
about a 1000 Jeans masses.  Bonnell, Bate, \& Vine (2003) presented
more realistic calculations with turbulent initial conditions,
but still isothermal and unmagnetized. While collisions were no longer
important for massive star formation, it was claimed that close
dynamical interactions were common for these stars during their
formation.

\vspace{-0.3in}
\section{Unraveling Orion}
\vspace{-0.1in}

\begin{figure}
\vspace{-0.03in}
\label{fig:schematic}
\plotfiddle{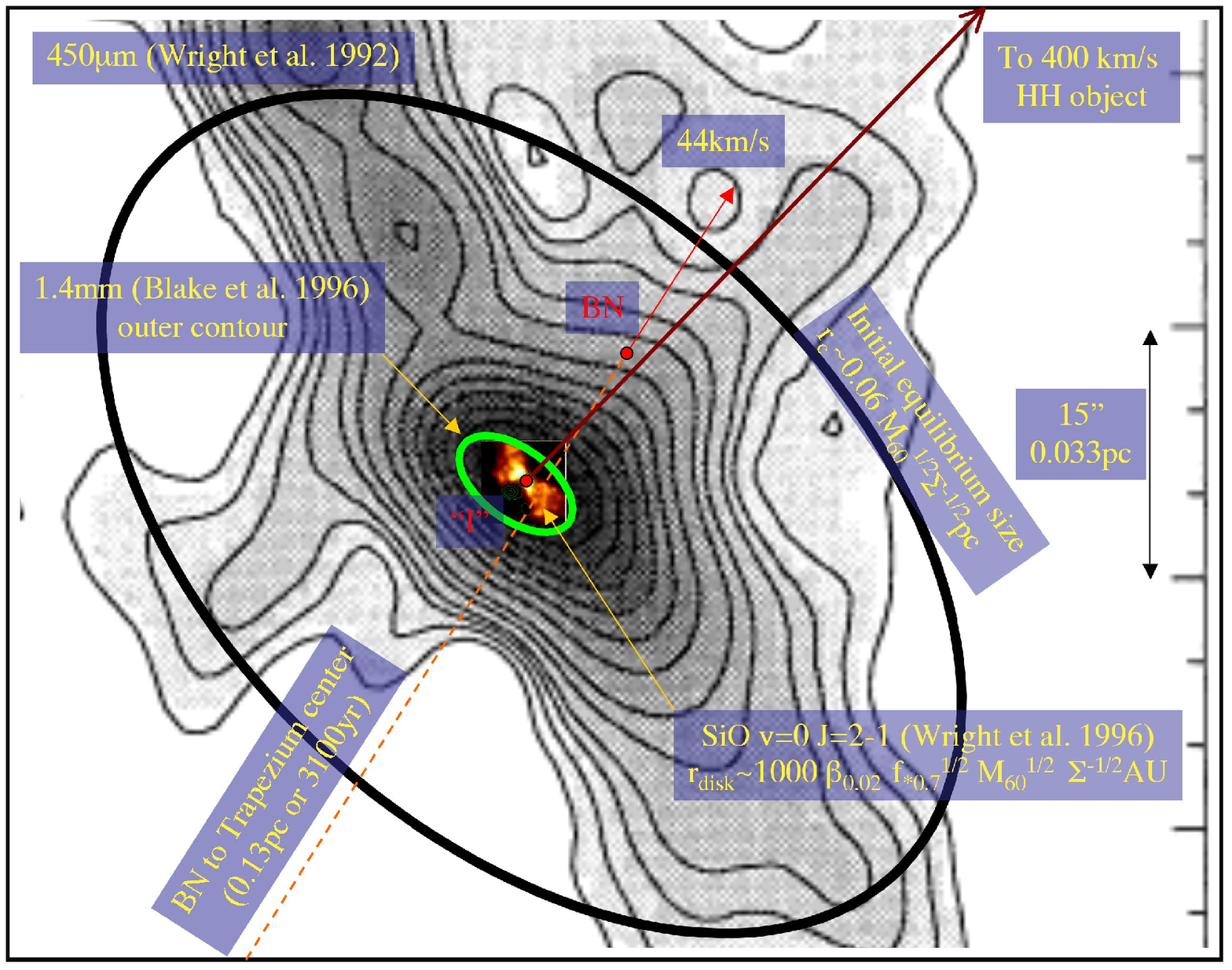}{3.8in}{0}{76}{76}{-222}{-35}
\caption{Schematic of Orion Hot Core region.}
\vspace{-0.33in}
\end{figure}

We focus on the Orion Hot Core region, which includes the
Kleinmann-Low Nebula (Fig. 1), $\sim 0.1$~pc NW in projection from the
Trapezium stars at the heart of the Orion Nebula Cluster (ONC), about
450~pc distant. The Hot Core region has a total luminosity $\sim
10^5\:L_\odot$ (Gezari, Werner, \& Backman 1998, hereafter G98). The
brightest 2$\rm \mu m$ source is the Becklin-Neugebauer (BN) object,
which is also a thermal radio source.  A second thermal radio source,
``I'', is located close to the center of the densest region of gas, as
traced by depth of the silicate absorption (G98). A weaker radio
source, ``n'', is nearby (Menten \& Reid 1995) but does not appear to
be particularly luminous (G98).

X-ray observations can provide a census of lower mass stars (Garmire
et al. 2000), with a $\sim 50$~ks Chandra observation sensitive to
X-rays from protostellar sources down to $A_V\sim60$ ($N_{\rm H}\sim
10^{23}\:{\rm cm^{-2}}$). Emission is detected 500~AU NW of BN and
coincident with ``n'', but not from ``I''. Including ``n'', there are
9 sources within 10\arcsec\ of source ``I'', all S and E of the NE-SW
axis. Only two do not have optical or near IR counterparts, and these
are both about 9\arcsec\ away. Taking a stellar surface density of
$4000\:{\rm pc^{-2}}$ for the larger scale distribution of stars
(Hillenbrand \& Hartmann 1998), we expect about 6 sources in
this region, so the stellar density is somewhat enhanced. If half of
the 9 sources are really within 4500~AU of ``I'', then the local
stellar density is $10^5\:{\rm pc^{-3}}$. This estimate is uncertain
because of Poisson statistics and source incompleteness, but 
is far below the level at which stellar
collisions are important.

Scoville et al. (1983) noted that the systemic velocity of BN and its
nebula was $+21\kms$, i.e. $+12\kms$ relative to the molecular cloud.
Plambeck et al. (1995) measured the proper motion of BN with respect
to ``I'' with a 9 year baseline. From this and more recent data
(Plambeck, private communication), we estimate a velocity of about
42~$\kms$ in the plane of the sky towards position angle $-33\deg$,
i.e. directly away from the Trapezium and ``I''. This gives a total
velocity of $44\:\kms$. This motion may explain the 500~AU displacement
of the X-ray emission. BN would have made closest approach to ``I''
about 460~yr ago, and would have been at the center of the Trapezium
3100~yr ago. Together with its estimated luminosity of
2500-$10^4\:{\rm L_\odot}$ (B98), we conclude that BN is a runaway
B3-B4 (8-12$\sm$) star, perhaps ejected from the Trapezium region.
Close passage near ``I'' may have triggered enhanced accretion (and
outflow) several hundred years ago, depending on their line-of-sight
separation.

The polarization vectors of 3.8~$\rm \mu m$ emission suggest that a
spatially concentrated source near ``I'' is responsible for much
of the Hot Core region luminosity (Werner, Capps, \& Dinerstein 1983).
If this is a single protostar, then the luminosity implies
a mass of $\sim 20\sm$ (MT). We do not
expect star formation from a core to be 100\% efficient, so the
initial mass in the region of the core that has now collapsed was
probably larger, perhaps by a factor of two.  There is still a
comparable mass of gas around the protostar that is still infalling.
Thus as a working model for the system, we consider
an initially $60\sm$ core, about 2/3 of which has already
collapsed. The bounding ambient pressure is influenced by the
self-gravity of gas in the clump (that has partly formed
the $\sim 1000\sm$ ONC) and by the pressure from feedback from
the massive Trapezium stars. We shall consider the case that the
pressure is equivalent to that due to self-gravity in the
central region of a clump with $\Sigma=1\:{\rm g\:cm^{-2}}$.

First consider the size of the initial core: it is $\sim 0.06\:{\rm
  pc}$ (12,000~AU, 26\arcsec) in radius. In the models of MT,
the cores were somewhat flattened due to a component of large
scale magnetic field support. An outline of the theoretical core is
shown in Fig.~1, superposed with the 450~$\rm \mu m$ (8\arcsec)
map of Wright et al. (1992). The sizes are comparable, with the
observed core being about twice as concentrated, as
might be expected midway through its collapse.
When constructing core spectra, care should be taken that the
regions probed at different wavelengths probe the same scales: e.g.
the radial extent of the outer contour in the 1.4~mm map of
Blake et al. (1996) is only $\sim 1800$~AU from ``I''.

For $\beta=0.02$, the disk size at this stage in the collapse is about
1000~AU, comparable to the extent of emission in the 1.4~mm continuum
(Blake et al. (1996) and SiO (v=0; J=2-1) maser line (Wright et al.
1995).  Indeed, Wright et al. interpreted the maser emission as
tracing a disk. Comparing the spectra of the emission
peaks of approaching and receding sides, separated by about
1\arcsec, most emission is in the range $-6$ to
$+14\kms$. For an inclination angle of the disk rotation axis to our
sight line of $65\deg$ (below), the true velocity difference
is then about $22\kms$. The enclosed mass inside $\sim 200$~AU is
thus $\simeq30\sm$.

SiO (v=1) maser transitions probe conditions of
higher density ($n_{\rm H}\sim$ $10^{10\pm1}\:{\rm cm^{-3}}$) and temperature
($T\gtrsim 10^3\:{\rm K}$), and are seen on scales $\sim50$~AU
(e.g. Doeleman, Lonsdale, \& Pelkey 1999), and may trace an outflow
along the NW-SE axis (Greenhill et al. 1998; however, see
Greenhill et al. 2003).

Tan \& McKee (2003, hereafter TM) presented a model for the density
structure of the outflow from a massive
protostar, including a low-density cavity along the rotation
axis. With parameters $m_*=20\sm$ and $\mds=10^{-4}\smyr$,
the density near the cavity and within $\sim 100$~AU of
the star is $n_{\rm H}\sim10^{8-9}{\rm cm^{-3}}$. This gas is
heated and ionized by the protostar, which may be close to the main
sequence.
Thermal bremsstrahlung emission from this {\it
  outflow-confined \ion{H}{2} region} explains the observed radio
spectrum, and the derived ionizing luminosity is consistent with
protostellar evolution models. In fact ``I'' appears elongated
along the outflow axis at 22 and 43~GHz (Menten \& Reid, in prep.).

The outflow speed should be about the protostellar escape speed, and
in the model of TM is $\simeq 1000\kms$. The on-axis,
well-collimated part of the flow, with line-of-sight velocities of
$\simeq 400\kms$ (for inclination angle of 65\deg\ adopted by TM),
should reach quite far from ``I''.  Taylor et al. (1986)
observed $\sim 300-400\kms$ blue-shifted
O{\small\rm I}\relax ~from Herbig-Haro objects about 1'(0.13~pc)
to the NW. 
Note that these velocities are much greater than those
probed by current maser observations near ``I''. 
We predict that faster maser velocities will be found if the search
region is expanded to higher (blue and red shifted) velocities. In
this model, individual spots with proper motion of only $\sim 10\kms$
must be decoupled and almost stationary with respect
to the gas flow.

The outflow is collimated, but not as a narrow jet: each logarithmic
interval in angle from the axis (cavity excluded) delivers about equal
momentum (Matzner \& McKee 1999). This flow interacts with the
surrounding turbulent and clumpy gas core, reducing the collapse
efficiency and sweeping up molecular gas to a range of
slower velocities.  On scales of $\sim 10\arcsec$ from ``I'',
Stolovy et al. (1998) report clumpy $\rm H_2$ emission with line
widths up to $\sim \pm 100\kms$.  We estimate a total momentum
injection $p_w \simeq 4500 (m_*/20\sm)^{1.4}\sm\kms$.
Depending on the ambient density, the wind may have penetrated
quite far in the clump. Within a few tenths of a
parsec, dense gas traced by $\rm NH_3$ (Wiseman \& Ho 1998) lies
in filaments perhaps sculpted by an outflow
from ``I''.
In this case there should be many gas clumps, stars, and protostars in
the flow.  The ``bullet+wake'' features (Allen \& Burton 1993) are
hard to understand in this context. They may be related to the
development of a thin-shell instability where the outflow has swept up
relatively uniform material (Stone et al. 1995). The outflow rate may
be highly variable, e.g. close passage of BN to ``I'' may have tidally
triggered enhanced accretion and outflow several hundred years ago.

\vspace{-0.1in}
\section{Conclusions}
\vspace{-0.05in}
An extension of low-mass star formation models, based on the collapse
of gas cores, to massive systems, can account in broad terms for many
of the observed features of the Orion Hot Core region, including the
core size, luminosity, accretion disk, compact \ion{H}{2} region, and
fast outflow. The situation is somewhat complicated by the presence of
other stars in the ONC: e.g. tidal interactions with other stars, such
as BN, may have led to sporadic enhancements in the accretion and
outflow. However, most stars are of low mass and have little influence
on the collapse.  Also the formation of the ONC is relatively
advanced, and these effects would have been less common for massive
stars forming in the earlier stages. We conclude that the Orion Hot Core
protostar provides good evidence in support of massive star formation
from coherent gas cores that become gravitationally unstable at
relatively large masses. These cores are rare both in terms of their
number and in the fraction of the total clump mass
they contain. Numerical simulation of the details of their
formation from a turbulent, magnetized, and self-gravitating medium is
an important goal.
\acknowledgments 
I thank many colleagues for helpful input, including Frank, Dave, and particularly Chris --- Happy Birthday! I am supported by a Spitzer-Cotsen
fellowship from Princeton and NASA grant NAG5-10811.

\end{document}